\documentclass[sigconf]{acmart}
\setcopyright{none}
\renewcommand\footnotetextcopyrightpermission[1]{} 
\usepackage{amsmath}
\usepackage{subcaption}
\usepackage{bm}
\newtheorem{assumption}{Assumption}

\newcommand{\RR}{\mathbb{R}}

\newcommand{\bZ}{\bm Z}
\newcommand{\bz}{\bm z}

\newcommand{\EE}{\mathbb E}

\newcommand{\NN}{\mathbb N}
\begin{document}
\pagestyle{plain}
\title{Treatment Effect Estimation Amidst Dynamic Network Interference in Online Gaming Experiments 
}

\author{Yu Zhu}
\affiliation{%
  \institution{University of California, Santa Cruz}
  \city{Santa Cruz}
  \state{CA}
  \country{USA}
}
\email{yzhu201@ucsc.edu}

\author{Zehang Richard Li}
\affiliation{%
  \institution{University of California, Santa Cruz}
\city{Santa Cruz}
  \state{CA}
  \country{USA}
}
\email{lizehang@ucsc.edu}

\author{Yang Su}
\affiliation{%
  \institution{Tencent}
 \city{Palo Alto}
  \state{CA}
  \country{USA}
}
\email{yaangsu@global.tencent.com}

\author{Zhenyu Zhao}
\affiliation{%
  \institution{Tencent}
   \city{Palo Alto}
  \state{CA}
  \country{USA}
}
\email{zzy287@gmail.com}

\renewcommand{\shortauthors}{Zhu, et al.}

\pagestyle{plain}
\begin{abstract}
The evolving landscape of online multiplayer gaming presents unique challenges in assessing the causal impacts of game features. Traditional A/B testing methodologies fall short due to complex player interactions, leading to violations of fundamental assumptions like the Stable Unit Treatment Value Assumption (SUTVA). 
Unlike traditional social networks with stable and long-term connections, networks in online games are often dynamic and short-lived. Players are temporarily teamed up for the duration of a game, forming transient networks that dissolve once the game ends. This fleeting nature of interactions presents a new challenge compared with running experiments in a stable social network. 
This study introduces a novel framework for treatment effect estimation in online gaming environments, considering the dynamic and ephemeral network interference that occurs among players. We propose an innovative estimator tailored for scenarios where a completely randomized experimental design is implemented without explicit knowledge of network structures. Notably, our method facilitates post-hoc interference adjustment on experimental data, significantly reducing the complexities and costs associated with intricate experimental designs and randomization strategies. The proposed framework stands out for its ability to accommodate varying levels of interference, thereby yielding more accurate and robust estimations. Through comprehensive simulations set against a variety of interference scenarios, along with empirical validation using real-world data from a mobile gaming environment, we demonstrate the efficacy of our approach. This study represents a pioneering effort in exploring causal inference in user-randomized experiments impacted by dynamic network effects.
\end{abstract}



\maketitle
\pagestyle{plain}

\section{Introduction}
A/B testing is heavily used to drive product decisions in the dynamic landscape of technology and digital platforms. Central to the effectiveness of these tests is the Stable Unit Treatment Value Assumption (SUTVA) \cite{Rubin1980}, which posits that the outcome of any unit (e.g., a user) in an experiment is unaffected by the treatment assignment of other units. This assumption is critical for unbiased estimation of treatment effects. However, in many real-world applications of A/B testing, particularly within online gaming, the SUTVA assumption is commonly violated. 

Online games, such as the multiplayer online battle arena (MOBA) game, frequently set up strategies to match players into teams. Connections and interactions between players are temporarily formed for the span of a single game session. Players' experiences are inherently affected by teammates and adversaries. For instance, consider an online experiment to assess the impact of game difficulty -- a scenario where players in the treatment group are assigned a less challenging version of the game. The assignment to treatment and control groups is delineated before the game starts. However, once the game session unfolds, the game difficulty level may be re-defined based on all the players' treatment assignments. In such a team-oriented environment, the exposure of a single player to a lower difficulty setting can ripple through the entire team, altering the actual treatment receipt of all involved and potentially affecting the game's outcome. And since the data are collected afterwards, interference among players becomes an inevitable issue. Furthermore, the network structure within online gaming is inherently dynamic and short-lived, continually evolving as game sessions conclude and new ones unfold. Each game session reshapes the network connections as players disperse and regroup. During the experiment period, players may enter multiple games and receive different numbers  of in-game treatments. This post-hoc network interference, stemming from the interactions between players and complicated by the number of treatments players received, may lead to biased estimations of causal effects. 


Recently, various experimental designs have been proposed to address the challenges posed by network interference, particularly in terms of mitigating spillover effects and enhancing the accuracy of treatment effect estimation. Notably, cluster-based randomized experiments \cite[e.g.,][]{Aronow2013, Eckles2016, Ugander2013, Walker2014}, have been at the forefront of these developments. These experiments group subjects into clusters based on their network connections, and then randomize the treatment at the cluster level rather than the individual level. Multi-level experiments \cite{Hudgens2008} introduce sophisticated frameworks where interventions are administered at multiple hierarchical levels within the network. Additionally, mixed experiments, as explored by \cite{Karrer2020}, simultaneously employ both unit-level and cluster-level randomization to capture a broader range of network effects.

Despite the innovative nature of these online experimental designs, there can be methodological, ethical, and cost-related challenges in applying these designs in the online gaming environment. For instance, it is not realistic to completely isolate players into distinct clusters without affecting their typical gaming experience. Players often interact in a dynamic, interconnected online space, making it difficult to create clear-cut, isolated clusters for experimentation. Additionally, cluster-based or multi-level experiments are typically inefficient to implement both computationally and financially.


More recent studies have focused on causal effect estimations under interference. One line of research particularly emphasized the concept of partial interference, which assumed that network interference is restricted within the non-overlapped groups \cite[e.g.,][]{Sobel2006, Tchetgen2012OnCI, Liu2014, Basse2017}. Another line of work studied the causal estimation in a more generalized setting, with arbitrary interference or introducing specific network graphs as extra information into the models \cite[e.g.,][]{Bowers2013, Manski2013, Goldsmith2013, Forastiere2016}. Some other papers discussed causal effect estimations under potentially mis-specified or unknown network interference \cite[e.g.,][]{Halloran2016, Choi2016, Forastiere2016}. Complementing these studies, \cite{Karwa2018} presented an innovative approach for causal estimations by developing a semi-parametric form of potential outcome. The potential outcome function includes the specification of the exposure mapping with network neighborhood information. \cite{Sävje2017} investigated the large sample properties of generalized treatment effect estimation under the unknown interference structure. 

While existing literature offers valuable insights into causal inference in the presence of network interference, treatment effect estimation in dynamic environments like online gaming remains less explored and comes with its set of unique challenges.
Specifically, within a short period, players repeatedly start new gaming sessions with potentially different team members. In contrast to applications like traditional social networks with a static and long-term network structure \cite[e.g.,][]{Xu2015, Gui2015}, the mechanisms of interference are not well-defined or consistent over time, thus all players can be subject to treatment spillover through a random process that experiment designers cannot control. Nevertheless, we will adopt similar framework as some of the previous work and define a more general exposure mapping in such settings.

Causal analysis within the literature on online gaming has been sparse. \cite{Meng2019} proposed a general causal analysis framework `exCause' for the real-time game sessions. \cite{He2021} employed causal inference techniques, specifically focusing on the estimation of heterogeneous treatment effects, to assess the impact of software updates and patches in games. \cite{Dong2020} innovated the novel Attention Neural Networks to refine the estimation of causal effects using mobile gaming data. However, it is notable that the network interference problem for causal inference, specifically in the online gaming area, remains unexplored.

In this study, we introduce an innovative framework that combines causal inference under network interference with the context of online gaming. Our main contributions can be summarized as follows: 
\begin{itemize}
    \item We formalize the problem of treatment effect estimation in online gaming with post-hoc network interference where the networks are dynamic and ephemeral. 
    \item We develop an estimator for treatment effect with interference, specifically when the completely randomized experimental design is required and the network structure is not explicitly known. 
    \item We evaluate the proposed estimator through a series of simulation studies with various interference settings, and demonstrate the accuracy and robustness of our approach, especially in comparison to more naive methodologies.
    \item We validate the effectiveness of the proposed framework through its deployment on real-world data from a mobile gaming online experiment.
\end{itemize}

The rest of the paper is organized as follows. In Section 2, we introduce the experimentation process in online mobile games, and discuss the limitations of the naive estimator for the Average Treatment Effect (ATE). In Section 3, we define the specific causal estimand of interest and describe our proposed framework for estimation in detail. In Section 4, we conduct a series of simulation studies under various interference scenarios and compare the performance of different estimators. Section 5 includes the application and validation of our proposed estimator using data from a real-world mobile gaming experiment, thereby demonstrating its practical efficacy.

\section{Online Experimentation of the Mobile Game}
Mobile gaming has seen unprecedented growth in recent years, becoming a significant part of the digital entertainment industry. Specifically, MOBA games have not only attained immense popularity but also cultivated a vast, global player community, largely due to their captivating gameplay mechanics and dynamic player interaction. Take a 5v5 game as an example, 5 players work together as a team to achieve objectives and defeat the opposing team. In this study, we ignore the influence of the opposite team and treat the games as the `Player versus Environment' (PVE).

In this study, we focus on the causal analysis of the `treated game' feature, an intervention designed to enhance player engagement and promote a more active gaming community. 
Specifically, when the `treated game' is activated in a game session, it modulates the game's experience level for the players. In an ideal scenario without interference among players, those in the treatment group would consistently participate in the `treated game', while those in the control group would never engage in the `treated game'.

Our experimental design incorporates a substantial segment of the online traffic, utilizing 40\% of it to gauge the effects of our treatment. This allocation is divided evenly, with 20\% of users randomly assigned to the treatment group and the remaining 20\% serving as a control group for a comparative analysis. This process can be seen as the initial stage of the randomized treatment assignment mechanism. 

\begin{figure}[h]
    \centering
    \includegraphics[width=\linewidth]{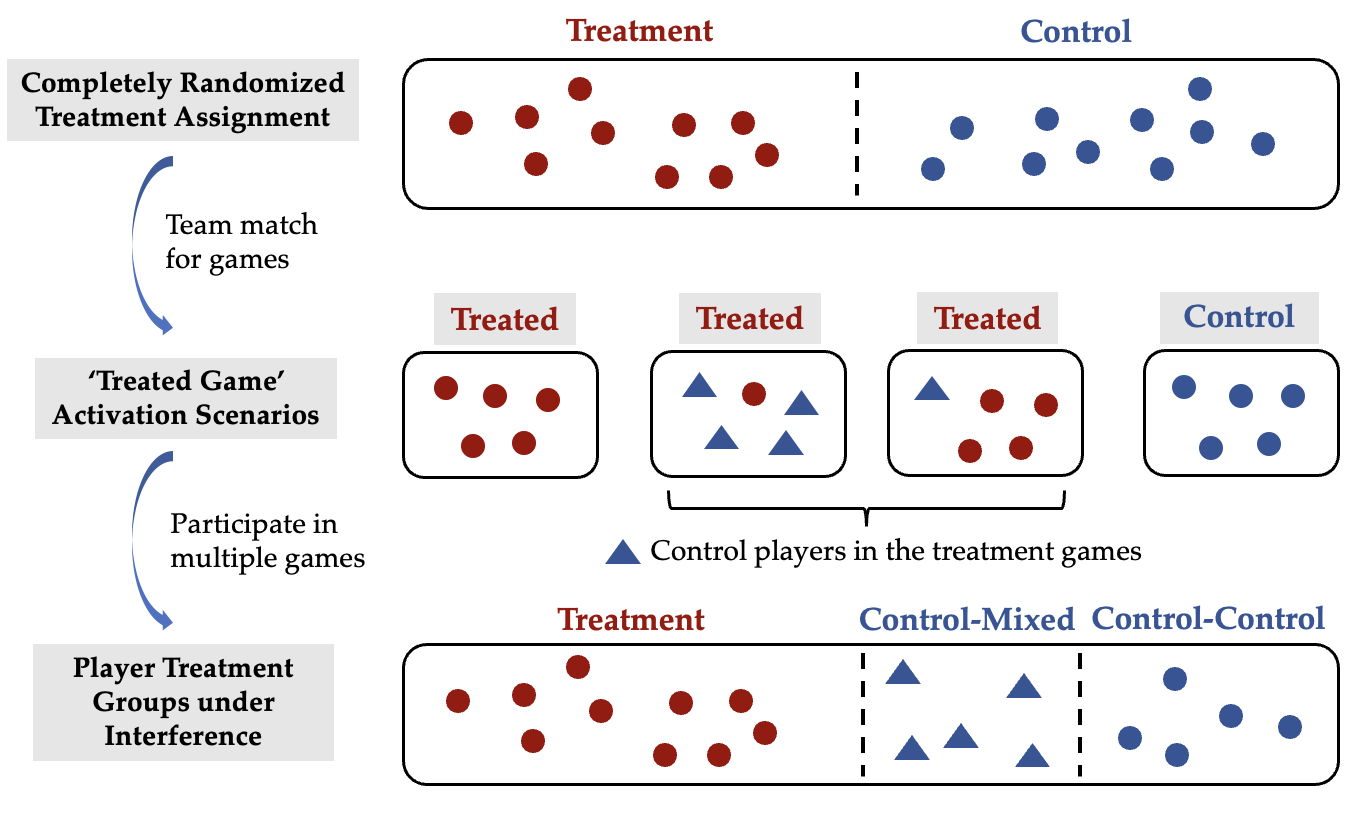}
    \caption{An illustration of the randomized experiment process with network interference during online games.}
    \Description{An illustration of the randomized experiment process with network interference during online games.}
    \label{fig:experiment_illustration}
\end{figure}

In the second stage, the players engage in games with an unknown team-matching process. The actual treatment status, or in other words, the activation status of the `treated game', for each player in a given game is determined based on a specific criterion:
all team members will receive the treatment if at least one member was initially designated to receive the treatment. Under this criterion, players who were initially assigned to the treatment group will consistently receive the treatment throughout the experiment. In contrast, those initially designated as `control' may experience a shift in experience due to the influence of their teammates from the treatment group. During the experimentation period, each player may participate in multiple gaming sessions. An illustration of this process is shown in Figure \ref{fig:experiment_illustration}.

Under this setting, players can be further categorized into three distinct groups based on their original treatment assignment and their engagement with the `treated game' feature:
\begin{itemize}
    \item \textbf{Treatment Group}: Players with an initial treatment assignment who consistently participate in the `treated game';
    \item \textbf{Control-Mixed Group}: Players originally assigned to the control group who have played both the standard (control) game and the `treated game';
    \item \textbf{Control-Control Group}: Players initially assigned to the control group, who exclusively play the control game without any exposure to the `treated game'.
\end{itemize}


In the rest of the paper, we adopt the following notation. For each player $i \in \{1, ..., N\}$, let $Z_i \in \{0, 1\}$ be the initial treatment assignment. We let $M_i \in \NN$ denote the number of treated games played by player $i$ and $Y_i \in \RR^+$ is the outcome of interest. As for the three treatment receipt groups, we use $T \subset \{1, ..., N\}$ to denote the set of players in the treatment group; $C_1 \subset \{1, ..., N\}$ the player in the control-mixed group and $C_0 \subset \{1, ..., N\}$ as the player in the control-control group. We let $X_i \in \mathcal{X}$ denote any pre-experiment covariates we have for each player.

\subsection{The Naive Estimator}\label{sec:Naive_Estimator}
The goal of this experimentation is to estimate ATE to validate whether this `treated game' intervention can significantly bring positive influence on user engagement. The ATE is commonly estimated using the strategy of Difference-in-Means (DiM) \cite[e.g.,][]{Karwa2018, Neyman1990, Neyman1935, Ding2017}, where the difference in the average outcomes between the treatment and control groups are calculated. 

If we ignore the post-hoc network interference and directly apply the DiM estimator to the treatment and control group, a naive estimator of the overall treatment effect is 


\begin{align}
    \hat{\tau}^{\mbox{naive}}  
    &= \frac{1}{|T|} \sum_{t \in T} Y_t^{obs} - \frac{1}{|C_0| + |C_1|} \sum_{c \in C_0 \cup C_1} Y_c^{obs} 
\end{align}

This naive estimator ignored the treatment received by players in the control group through treated games. This results in an over-estimation of the baseline effect based on the control group data. Consequently, such over-estimation introduces bias into our analysis, leading to under-estimation of treatment effect. Similarly, one might define the naive estimator by replacing the second term as the average outcome in the control-control group. Such estimators are also biased since intrinsically, the control-control group tends to include players who are less active, as opposed to those in the control-mixed group, who are generally more active and thus have a higher chance of participating in the `treated game'. We demonstrate such biases in Section 4 and 5.

\section{Network Experiment Analysis}
In this section, we move beyond the naive estimators and formally define the causal estimand of interest in our setting, together with assumptions that allow us to define a valid estimator of the average treatment effect.

\subsection{Causal Estimand}
We follow the potential outcome framework and use $Y_i(\bZ)$ to denote the outcome that would be observed had the treatment status for all players been set to the vector $\bZ$. The potential outcome is indexed by the treatment status of all players as we allow arbitrary interference through treated games. This potential outcome is well-defined if the following assumption holds.

\begin{assumption}[No multiple treatment]\label{as:1}
If $\bZ = \bz$, then $Y_i = Y_i(\bz)$.
\end{assumption}

Specific to our case, as the number of treated games played by each player is known, it can serve as the so-called `exposure mapping' in the interference literature \citep{aronow2017causal}, i.e., we can assume that treatment of other players only affects unit $i$ through the number of treated games that player $i$ played. 

\begin{assumption}[Exposure mapping]\label{as:2}
Let $\bZ = (Z_i, \bZ_{-i})$. There exists a exposure function $g : \{0, 1\}^{N} \rightarrow \NN$ such that the following equality holds
\[
Y_i(Z_i, \bZ_{-i}) = Y_i(Z_i, \bZ_{-i}') 
\]
if $g(Z_i, \bZ_{-i}) = g(Z_i, \bZ_{-i}')$. We further assume that $M_i = g(Z_i, \bZ_{-i})$, so the potential outcome can be simplified into $Y_i(m, z)$.
\end{assumption}


We can define the causal effect for a fixed level of exposure $m > 0$, 
\begin{align}
    \tau(m) &= \frac{1}{N}\sum_{i = 1}^N \EE\left[Y_i(M_i = m, Z_i = 1) - Y_i(M_i = 0, Z_i = 0)\right].
\end{align}

We refer to $\tau(m)$ as the average treatment effect in the rest of the paper as it measures the average effect on the players' potential outcomes when they experience certain number of `treated games', compared to when they don't experience the treated game at all. This causal analysis helps in understanding how varying intensities or types of treatment influence players' behavior or game experience. 

Another estimand that is of interest is the overall effect $\tau$, defined as the ATEs weighted by players' natural distribution of the number of games they engage in with the `treated game' feature activated in the population. Since the initial treatment assignment is randomized, we have
\[
\tau = \sum_{m > 0} \tau(m) P(M_i = m \mid Z_i = 1).
\]

\subsection{Estimation}

Since the exposure received by each player cannot be randomized, we further make the common unconfoundedness assumption for both the initial treatment assignment and exposure as follows.

\begin{assumption}[Unconfoundedness of the joint treatment]\label{as:3}
\[
Y_i(m, z) \perp Z_i, M_i \mid X_i 
\]
\end{assumption}

Under Assumptions \ref{as:1} to \ref{as:3}, we have the following identification result:
\begin{align}
 \EE\left[Y_i(m, z)\right] 
 &\overset{(a)}{=}  \EE_{X}\left[\EE\left(Y_i(m, z) \mid X_i\right)\right] \\
 &\overset{(b)}{=} \EE_{X}\left[\EE\left(Y_i(m, z) \mid X_i, M_i = m, Z_i = z\right)\right] \\
 &\overset{(c)}{=}  \EE_{X}\left[\EE\left(Y_i \mid X_i, M_i = m, Z_i = z\right)\right] 
 \end{align}
Here, the equation $(a)$ is due to the law of total expectation; $(b)$ is due to Assumption \ref{as:3} and $(c)$ is due to Assumption \ref{as:1} and \ref{as:2}.

For the estimand of interest, we only need to evaluate $\EE\left[Y_i(m, 1)\right]$ and $\EE\left[Y_i(0, 0)\right]$. For the latter, we have pre-experiment data for all the players in the study, which can provide robust estimation for the outcome function without intervention. We denote the predicted outcome from this model to be
\[
\hat{\mu}(x) = \hat\EE\left[Y \mid Z = 0, M = 0, X = x \right].
\]
For the treated players, the level of exposure, $m$, can have a highly unbalanced distribution. Thus models that directly estimate the outcome $\EE\left[Y_i \mid X_i, M_i = m, Z_i = 1\right]$ can be unstable and computationally challenging.
Therefore, we propose a weighted estimator of $\EE\left[Y_i(m, 1)\right]$ using Inverse Probability Weighting (IPW) \cite[e.g.,][]{ROSENBAUM1983, Hirano2001}. IPW is a statistical technique used to adjust for potential selection bias where random assignment is not possible. The core idea of IPW is to re-weight the data so that the weighted sample resembles a randomized experiment. That is, we adopt the following unbiased estimator \citep{Hajek1971} for $m > 0$,
\begin{equation} \label{eq:Ym}
\hat{\EE}\left[Y_i(m, 1)\right] 
= \left[\sum_{i=1}^N \frac{\bm{1}\{M_i = m\}}{\hat{e}_{m}(X_i)}\right]^{-1} \sum_{i=1}^N \left[ \frac{Y_i^{obs} \cdot \bm{1}\{M_i = m\}}{\hat{e}_{m}(X_i)} \right]
\end{equation}

where $\hat{e}_{m}(x) = P(M_i = m \mid X_i = x)$ is the estimated propensity score for being exposed to $m$ treated games. Putting the two estimators together, we have an unbiased estimator of the average treatment effect
\begin{align}
\hat{\tau}(m) &= \left[\sum_{i=1}^N \frac{\bm{1}\{M_i = m\}}{\hat{e}_{m}(X_i)}\right]^{-1}  \sum_{i=1}^N \left[\frac{Y_i^{obs} \cdot \bm{1}\{M_i = m\}}{\hat{e}_{m}(X_i)} \right] - \frac{1}{N} \sum_{i=1}^N \hat{\mu}(X_i)
\end{align}

In practice, observations with large $m$ can be extremely sparse. One common strategy is to truncate the number of treated games and treat all $m$ above a certain threshold to be a single category. The estimation of propensity scores for multi-level treatment is notably more complex than in binary cases and requires more flexible models to avoid unstable or extreme weights in the IPW estimator.
We estimate the propensity score function $\hat e_m(x)$ with flexible predictive models, as described in the Section 4 and 5. 

\section{Simulations}
In this section, we evaluate the proposed framework with simulation studies reflecting different scenarios of network interference. The source codes are provided in this link\footnotemark{} \footnotetext{\url{https://github.com/YuZoeyZhu/-KDD-2024-Network-Interference-Online-Gaming/tree/main}}.
We assume a 5v5 game setting, i.e., $5$ players participate in each game. Let the total number of players $N = 1000$. Under the completely randomized treatment assignment, we generate the treatment assignment $Z_i \sim Bernoulli(0.5)$ for each player $i$. 
For each player, we also generate a single pre-experiment covariate $X_i \sim Beta(0.5, 0.5)$. The player-matching process is simulated as follows. 
For the $j_{th}$ round of the game,
\begin{itemize}
    \item Sample the number of treatment players $n_{T(j)} \in \{0, 1, ..., 5\}$ based on the pre-set probability. The number of control players $n_{C(j)} = 5 - n_{T(j)}$;
    \item Sample $n_{T(j)}$ players from the treatment group with probability $p(i \mid Z_i = 1) \propto \frac{0.8}{|T|} + 0.2(\frac{X_i}{\sum X_i})^2$; 
    \item Sample $n_{C(j)}$ players from the control group with probability
    \begin{align}\nonumber
    p(i \mid Z_i = 0) &\propto \begin{cases}
        X_i \bm{1}\{X_i < 0.2\} & \text{if } n_{T(j)} =  0,\\
        \frac{0.8}{|C_0| + |C_1|} + 0.2\left(\frac{X_i}{\sum X_i}\right)^2  & \text{ o.w.}
    \end{cases}
    \end{align}
\end{itemize}
We generate $N_g$ rounds of game. Let $M_i$ denote the number of `treated games' player $i$ participated.
We assume the potential outcome follows the following distribution with mean depending on both the feature $X$ and number of treatment games $M$:
\begin{align}\nonumber
    Y_i \mid M, X &\sim Exponential(1/\lambda)\\\nonumber
    \lambda &= 0.5M^{1/2} + 2X + 0.5X\bm{1}\{X > 0.5\} +  0.5M^{1/2}X
\end{align}
This synthetic data generating process mimics the distribution of real data where the heterogeneity of $Y$ increase for larger $X$. For example, if $X$ measures the level of activity of a player, active players tend to have larger variance in $Y$ than less active players.  

\begin{figure*}
    \centering
    \includegraphics[width=\textwidth]{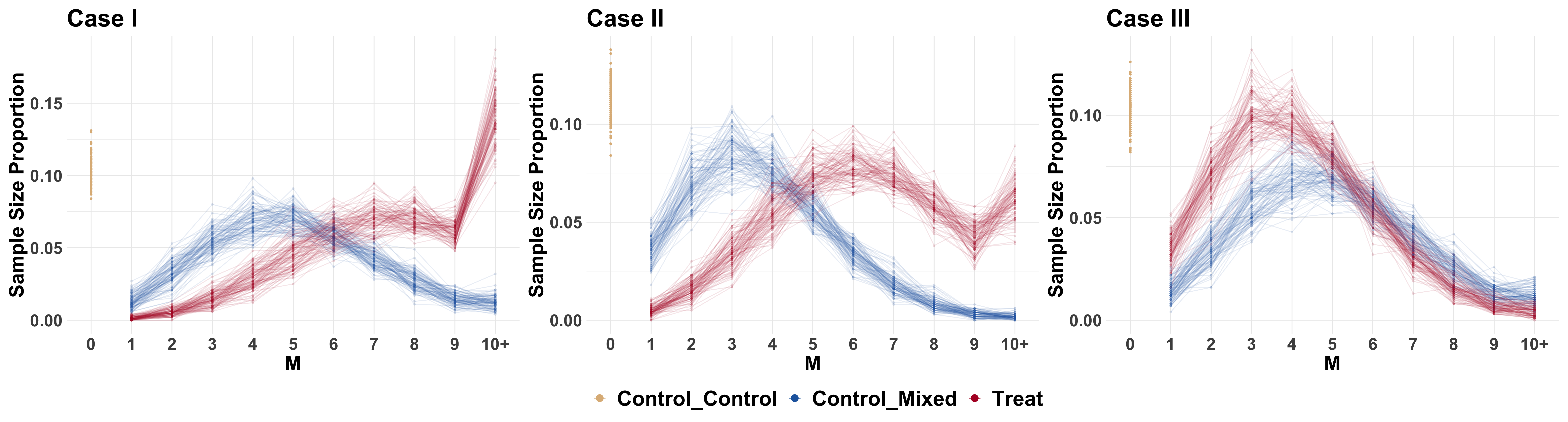}
    \caption{Visualization of the sample size proportions versus different treatment levels $M$ for the treatment (in red lines), control-mixed (in blue lines) and control-control (in yellow dots) groups with 50 simulated data sets under each simulation setting. It shows the different interference structures under different treatment player matching probabilities and number of game generations.}
    \label{fig:simulation_setup}
\end{figure*}

The simulation setting allows us to control the amount of interference by varying the probability of sampling players in the treatment group. More specifically, we consider three different settings with a progressively increasing level of interference in the control-mixed group from Case I to Case III.
The number of games $N_g$ and the distribution of exposure $p(M)$ are summarized in Table \ref{tab:simulation_setup}. 
Figure \ref{fig:simulation_setup} illustrates the distribution of sample size proportions across different values of $M$ for all three groups of players: treated, control-mixed, and control-control group. The amount of interference, increasing from I to III, is the sample size proportion of control-mixed group under each $M$ relative to the treatment group, which can be compared via the overlapping area under the blue line (the control-mixed group) and red line (the treatment group).  In addition, in Case I, the sample size proportions are relatively small in both the treatment and control-mixed groups at lower $M$ levels, which poses more challenges in estimating $\tau(m)$ when $m$ is small. In Case II, the probability of $M$ in the control-mixed group is skewed to the right with the mode at 3. The sample size proportions are slightly smaller in both small and large $M$s, but is overall more balanced comparing with Case I. In Case III, the probability of $M$ in the control-mixed group and treatment group are both skewed to the right with similar modes. In contrast to Case I, the sample size proportions are small in both groups at higher $M$s. In all three cases, around $10\%$ of samples are in the control-control group.


In the simulation, we assume $\mu(x)$ is known, which is a reasonable assumption when there are enough pre-experiment data without treatment. Thus we focus on the estimation of the counterfactual under non-zero exposures. We truncate the $M$ at 10, treating $M$ as categorical with values in $\{0, 1, 2, ..., 9, 10+\}$ and apply the XGBoost classification model \cite{ChenTianqi2016} to estimate the propensity scores $\hat e_m(x)$. 
We also evaluate a modified version of the proposed estimator by using only the players in the treatment group to estimate $\hat{\EE}\left[Y_i(m, 1)\right]$ in equation \ref{eq:Ym}, which we refer to the `\textbf{proposed estimator without control-mixed}'.

Let $\bar Y^{(m)}$ be the average outcome for players in the treatment group who are exposed to $m$ treated games, $\bar Y^{\mbox{c}}$ and $\bar Y^{\mbox{cc}}$ be the average outcome for players in the control group and control-control group respectively. We compute our estimator to two naive estimators ignoring interference: 
\begin{itemize}
    \item \textbf{Naive}: This is the simplest DiM estimator ignoring interference and confounding and only computes the contrast between players in the treatment group who are exposed to $m$ treated games with the whole control group
    \begin{align}
    \hat{\tau}^{\mbox{naive, 1}}  
    &= \bar Y^{(m)} - \bar Y^{\mbox{c}}
    \end{align}

    \item \textbf{Naive - w/o Control-Mixed}: Alternatively, we may remove all players subject to interference and consider the comparison to the control-control group only, 
    \begin{align}
    \hat{\tau}^{\mbox{naive, 2}}  
    &= \bar Y^{(m)} - \bar Y^{\mbox{cc}}
    \end{align}
\end{itemize}

\begin{figure*}
    \centering
    \includegraphics[width=0.9\textwidth]{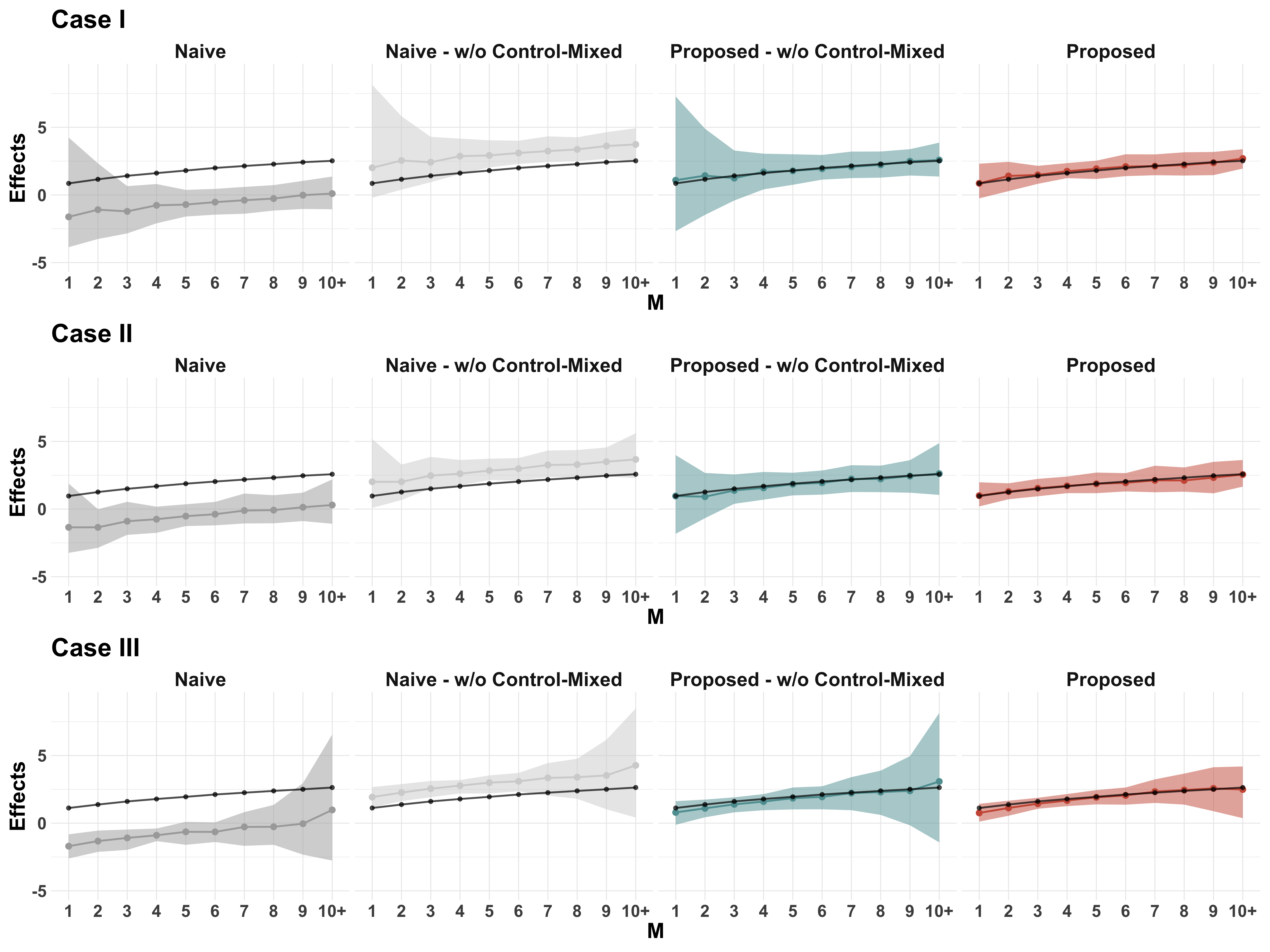}
    \caption{Comparisons of the Average Treatment Effect $\tau(m)$ estimations under each treatment level $M$ for the four estimators, with the corresponding mean and 95\% interval of the estimated effects under 100 simulated data sets in each simulation setting. The black line is the ground truth of $\hat{\tau}(m)$.}
    \label{fig:simulation_results}
\end{figure*}

To evaluate the performance of each estimator, we generate 100 datasets in each case and compare the mean and 95\% uncertainty interval of the estimated effects with the truth for $\hat{\tau}(m)$. As shown in Figure \ref{fig:simulation_results}, in all three cases, the first naive estimator tends to overestimate the effects in all levels of $M$, whereas the second naive estimator ignores the control-mixed group's consistently overestimated effect. This is as expected as in the former case, $\EE\left[Y(0, 0)\right]$ is over-estimated as some of the players in the control group are exposed to treatment through interference; whereas in the latter case, $\EE\left[Y(0, 0)\right]$ is under-estimated as the players in the control-control group are more likely to be less active players. Both versions of the proposed estimator show improvement over the naive estimators in general. We also observe that the proposed estimator without using the control-mixed group exhibits larger uncertainty in general, especially for levels of $M$ with a small sample size. By incorporating both the treatment and control-mixed group, the proposed estimator is able to achieve the smallest bias with low variation. 

This simulation study offers clear insights into the performance of our estimator across various probabilities of treatment player matching. In real cases, these matching probabilities can be strategically adjusted. For instance, selecting an appropriate matching probability can help maintain a balance between enhancing the user engagement of new features and improving the accuracy of the estimator. It can be further explored to manage and control network interference, thereby aligning the experimentation with practical business considerations.

This simulation study provides a straightforward insight of the performance of the estimator under different treatment player matching probability. In the real case, the matching probability can be adjusted on purpose with business concerns. And we discuss it as one of the strategies to control the interference.

\begin{table*}
        \centering
        \caption{Three different settings for the simulation.}
        \begin{tabular}{l|ccccccc}
         \toprule
         \toprule
                  Case   & $N_g$ & $p(M = 0)$ & $p(M = 1)$ & $p(M = 2)$ & $p(M = 3)$ & $p(M = 4)$ & $p(M = 5)$\\
                       \midrule
           I  & 2000 & 0.40 & 0.10 & 0.10 & 0.10 & 0.10 & 0.20\\
          II  & 1000 & 0.06 & 0.02 & 0.19 & 0.23 & 0.34 & 0.16\\
          III & 1000 & 0.20 & 0.34 & 0.07 & 0.16 & 0.06 & 0.17\\
           \bottomrule
        \end{tabular}
        
        \label{tab:simulation_setup}
\end{table*}

\section{Case Study}
In this case study, we consider a online experimentation for the `treated game' feature of a MOBA game from Tencent. We obtained the experiment dataset with a total of 58,565 players for a two-week experimentation period. And we collected a pre-experiment dataset from the same set of players, covering the two weeks immediately preceding the start of the experiment. The control-control group takes around 7.8\% of the whole sample.

In this experiment, we focus on monitoring a specific target metric (TM) for each player. This business metric is a measurement of a player's engagement in the mobile game. An increase in TM generally indicates a higher level of player involvement and satisfaction with the game, which can lead to increased loyalty, longer-term player retention, and potentially greater revenue through in-game purchases. We expect to see a significant increase in TM with the new `treated game' feature.

\begin{figure}[h]
    \centering
    \includegraphics[width=\linewidth]{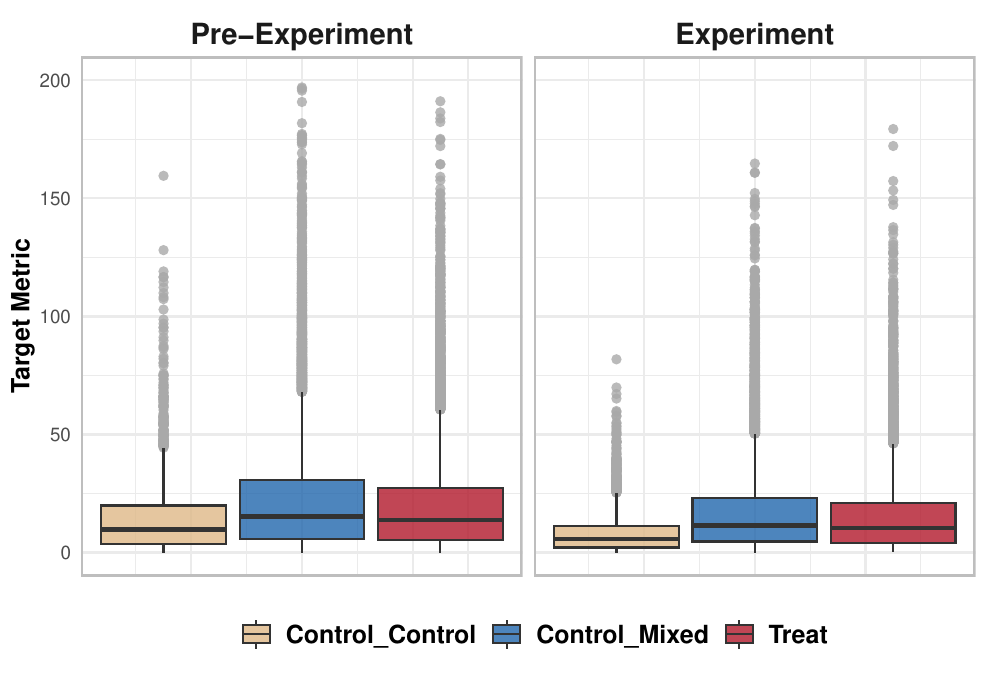}
    \caption{The side-by-side boxplots of the target metric (TM) for the control-control, control-mixed and treatment groups in the pre-experiment and experiment data sets.}
    \label{fig:EDA_TOT}
\end{figure}

This specific TM, being inherently positive, displays a distribution with decreasing tread and a long right tail. Figure \ref{fig:EDA_TOT} illustrates the distributions of TM across the three groups, comparing the experimental period and the pre-experimental phase data. In both datasets, the TM for the control-control group is relatively smaller, aligning with the inference that this group consists of less active players. Additionally, the pre-experiment TM generally surpasses the experiment TM, which could be attributed to various factors, such as in-game events, public holidays, etc.

The pre-experiment features, which serve as potential confounders, encompass a variety of descriptors that capture different characteristics of the players. These include metrics that measure players' gaming abilities, their levels within the game, and their in-game purchases.

As for the pre-processing of the data, we treat the observations with TM $\geq 60$ as outliers and remove them from the data. We also truncate $M$ with a threshold of 21. 

\subsection{Estimation of $\hat{\mu}(x)$}
We estimated $\hat\mu(x)$, the expected potential outcome under no treatment, as the sum of the observed TM during the pre-experiment period and a covariate-dependent increment during the experiment period. That is, we assume the difference of the potential outcomes between the pre-experiment and experiment periods under no treatment can be explained by the observed covariates. We estimate this difference with a linear model using the control-control group, and make predictions for the control-mixed group and treatment group players to get $\hat \mu(x)$. This approach is related to the Difference-in-Difference (DiD) estimator\cite{Ashenfelter1978, Ashenfelter1999, Bertrand2003}.
\subsection{Estimation of propensity score}
We used the XGBoost classification model to estimate the propensity score with $M \in \{0, 1, 2, ..., 20, 21+\}$. To implement the model, 
We used 5-fold cross-validation to find the optimal hyper-parameters, with learning rate $\eta = 0.3$ and the max depth = 6. The overall accuracy in the testing set is $78.5\%$. The propensity score is estimated with the predicted probability of the corresponding level of exposure.

\subsection{Performance}
\begin{figure}[h]
    \centering
    \includegraphics[width=\linewidth]{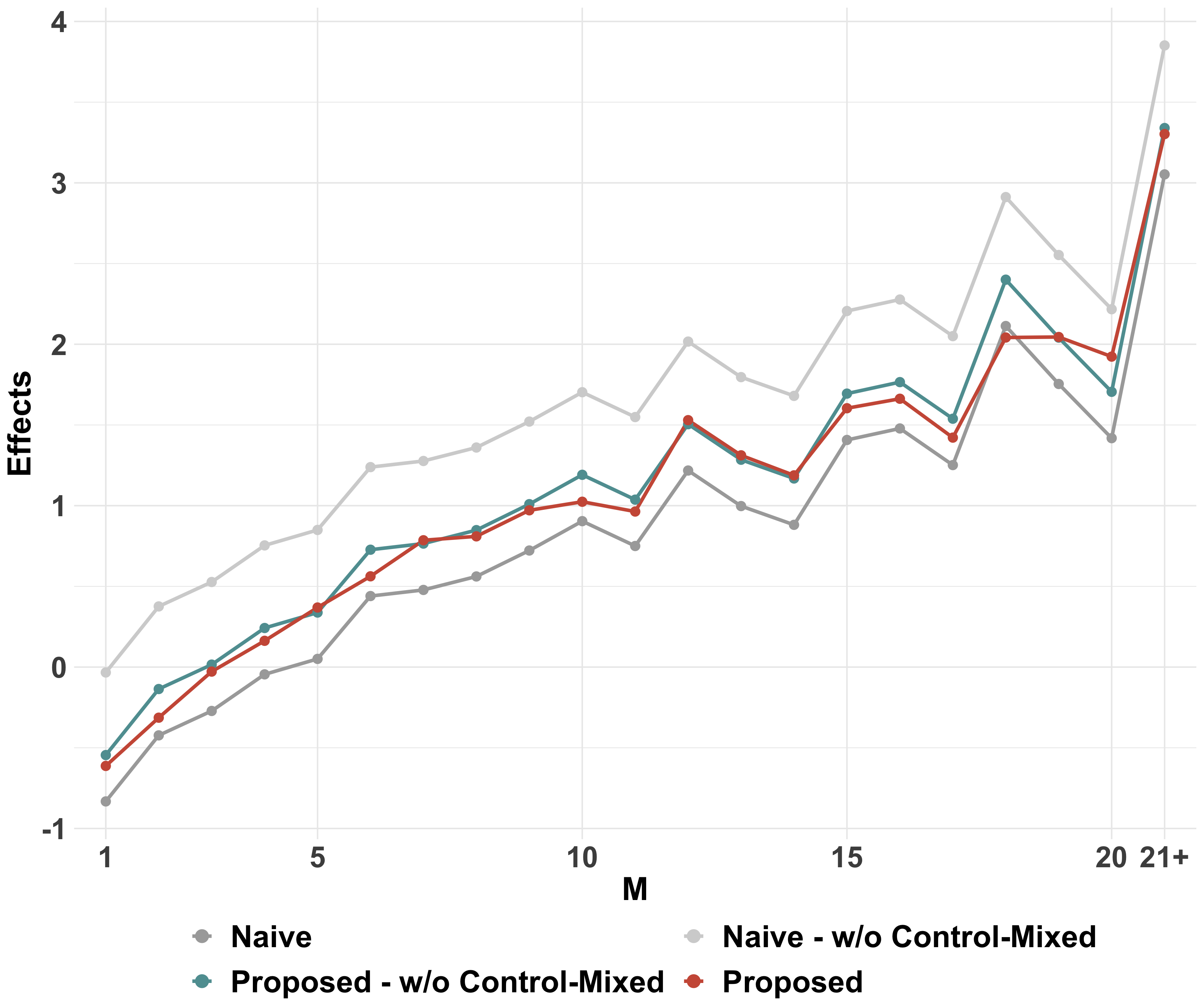}
    \caption{The Average Treatment Effect $\tau(m)$ estimations under each treatment level $m$ for the four estimators based on the real online gaming data set.}
    \label{fig:realdata_tau_M_comparison}
\end{figure}

Figure \ref{fig:realdata_tau_M_comparison} presents the effect estimations with the comparisons of four estimators across multiple treatment levels. The `Naive' estimator, represented by the dark grey line, suggested a conservative estimation for the treatment effects as expected. It also provides a baseline for comparison against more sophisticated approaches. The `Naive without Control-Mixed' estimator in the light grey line shows overall higher effects estimation compared with others, indicating that excluding the mixed-control data naively may lead to over-estimation. The proposed estimator is between the two naive approaches, and also close to the trajectory of the `Proposed without Control-Mixed' estimator.

The estimated treatment effects show an overall increasing trend as the treatment level $M$ grows, regardless of the estimation approaches. Besides, the increasing relationship is non-linear, with fluctuations and diminishing incremental benefits beyond a certain level of $M$. This is expected because a higher level of treatment is presumed to present a stronger influence on the outcome, and the treatment effects reach a plateau as the treatment level approaches a saturation point where additional increments in $M$ no longer yield proportional increases in the outcome. The relatively small sample sizes under large exposures also lead to the noiser estimations of the ATE.


Lastly, Table \ref{tab:tau} shows the estimation of overall treatment effect $\tau$ for all the estimators.

\begin{table}
  \caption{Marginal Treatment Effect $\hat{\tau}$}
  \label{tab:tau}
  \begin{tabular}{cc}
    \toprule
    Estimator & $\hat{\tau}$ \\
    \midrule
    \text{Naive} & 0.960 \\
    \text{Naive - w/o Control-Mixed} & 5.270 \\
    \text{Proposed - w/o Control-Mixed} & 2.509 \\
    \text{Proposed} & 2.113 \\
  \bottomrule
\end{tabular}
\end{table}

\section{Conclusion}
In conclusion, our comprehensive study provides a novel framework for analyzing causal effects in the context of online gaming, where traditional A/B testing faces the challenges of post-hoc network interference. By focusing on the `treated game' feature of the mobile MOBA game, we have demonstrated the potential of our proposed framework to accurately estimate treatment effects amidst the complex dynamics of player interactions.

It is worth noticing that certain limitations could impact the generalizability and applicability of our findings. First, our model relies on assumptions that may not hold across all gaming environments or user demographics. Moreover, our model presumes that the post-hoc network interference is uniformly distributed across all players, an assumption that might oversimplify the complex and often unique interactions between players. Besides, potential selection biases may not be fully accounted for due to the unknown team-matching strategy. For future work, we can further refine and validate our model with the underlying assumptions. For instance, exploring models that account for heterogeneity in player behavior or that explicitly model the unique network structures inherent in different gaming communities.

The potential applications of our work extend beyond the realm of online gaming. The principles and methodologies we have outlined could be adapted to other digital platforms and social networks where user engagement and interaction are pivotal to the product's success. Specifically, we can generalize our work to the exploration of dynamic network interference scenarios, such as those encountered in real-time strategy or ride-sharing platforms like Uber with continuously evolving user interactions.




\begin{acks}
We would like to thank Sizhe Zhang, Yifeng Huang, Junlong Zhou, Ke Nie and Ning Zhang for useful comments and discussions.
\end{acks}

\bibliographystyle{ACM-Reference-Format}
\bibliography{sample-base}


\end{document}